\documentclass[epj]{webofc}
\usepackage[varg]{txfonts}   % Web of Conferences font
\newcommand{\beq}{\begin{equation}}
\newcommand{\eeq}{\end{equation}}

\newcommand{\beqa}{\begin{eqnarray}}
\newcommand{\eeqa}{\end{eqnarray}}

\newcommand{\p}{\partial}

\wocname{EPJ Web of Conferences}
\woctitle{ICNFP 2016}
\begin{document}
\selectlanguage{english}
\title{Some Developments in Gribov's Approach to QCD}\footnote{5th International Conference on New Frontiers in Physics, 6-14 July, 2016, Orthodox Academy of Creta, Greece }
%
% subtitle (optional, strongly discouraged)
%
%%%\subtitle{Do you have a subtitle?\\ If so, write it here}

\author{Patrick Cooper\inst{1}\fnsep\thanks{\email{cooperp@duq.edu}} \and
        Daniel Zwanziger\inst{2,3}\fnsep\thanks{\email{dz2@nyu.edu}} %\and
        %Third author\inst{1,3}
        % etc.
}

\institute{Physics Department, Duquesne University, 600 Forbes Ave, Pittsburgh, PA 15282, USA
\and
           Physics Department, New York University, 4 Washington Place, New York, NY 10003, USA
\and
          Speaker
%\and
%           The last address here
}

\abstract{%
  We review several developments in the formulation of QCD provided by the GZ action.  These include the GZ-action at finite temperature, the relation of the horizon condition and the Kugo-Ojima confinement criterion, the relation of the horizon condition and the dual-Meisssner effect, the alternative derivation of the GZ action provided by the Maggiore-Schaden shift,  and the spontaneous breaking of BRST symmetry.  We conclude with a proposal for the definition of physical states in the presence of BRST breaking.
}
\maketitle
\section{Introduction}
\label{intro}
This article is dedicated to the memory of Vladimir Gribov.  In it we review several developments that emerged from his paper of 1977 \cite{Gribov:1977wm} on the quantization of gauge fields.  In it he explained how the non-perturbative nature of non-abelian gauge theory, manifested in what are now called ``Gribov copies." leads to a long-range, potentially confining force.

\section{No Confinement without Coulomb Confinement}

Gribov's insight into the mechanism of confinement is substantiated by the theorem \cite{Zwanziger:2002},
\beq
V_{\rm coul}(R) \geq V_{\rm wilson}(R)
\eeq
for $R \to \infty$, where the color-Coulomb potential is the
temporal gluon propagator in Colomb gauge,
\beq
D_{00}(R, T) = V_{\rm coul}(R) \ \delta(T) + \mathrm{non-instantaneous}.
\eeq
When the Wilson potential is linearly rising, the color-Coulomb potential is linear or super-linear.

%%%%%%%%%%%%%%%%%%%
\section{All horizons are one horizon}
%%%%%%%%%%%%%%%%%%%%

%%%%%%%%%%%%%%%%%%%%%%%%%%%%%%%%%%%%%%%%%%%%%%%%%%%  Figures for Dan %%%%%%%%%%%%%%%%%%%%%%%%%%
\begin{figure}
        \begin{center}
                \includegraphics[width=7cm,clip]{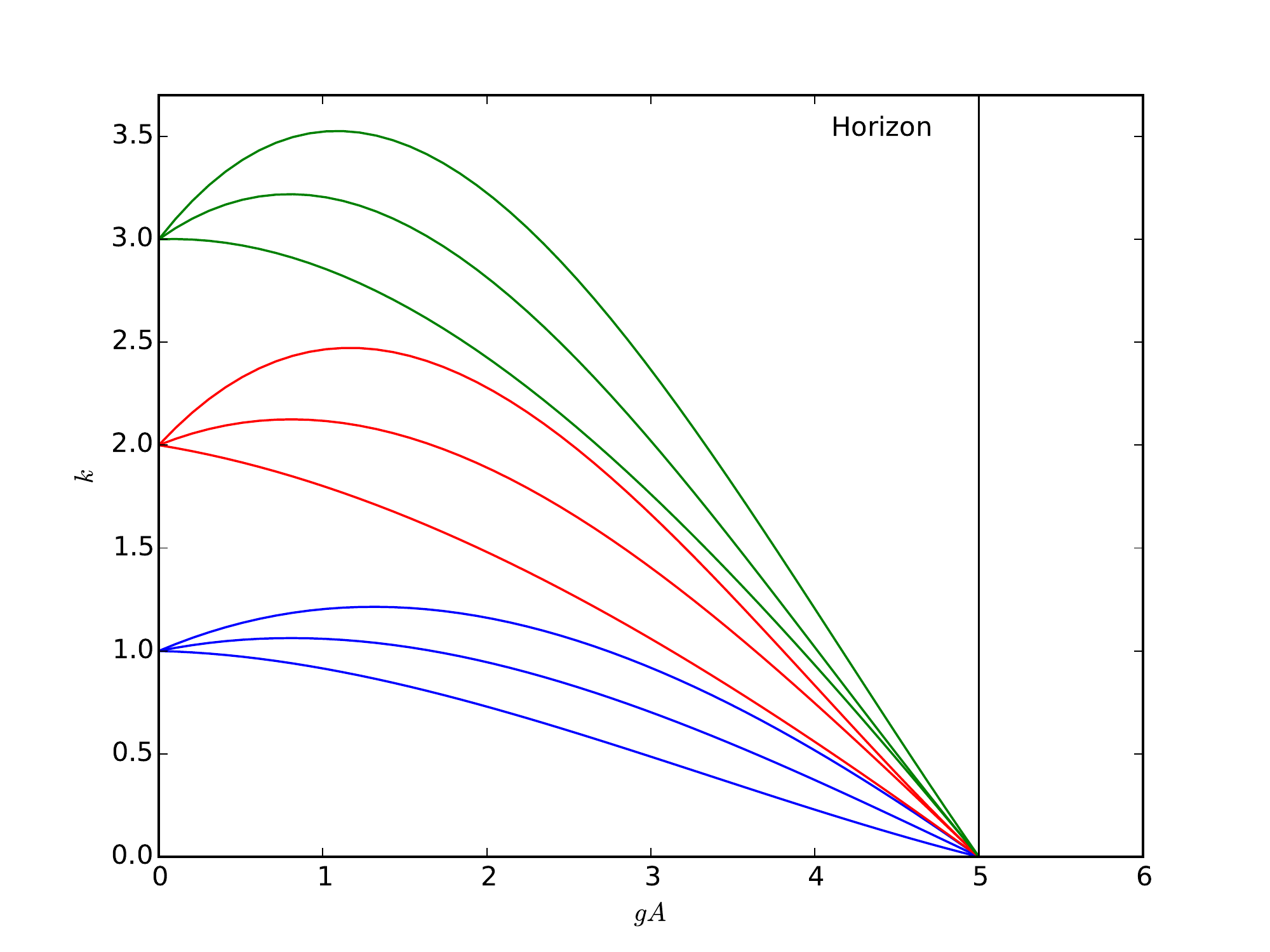}
                \caption{Cartoon of the eigenvalues of the Fadeev-Popov operator.}
                \label{plot}
        \end{center}
\end{figure}
%%%%%%%%%%%%%%%%%%%%%%%%%%%%%%%%%%%%%%%%%%%%%%%%%%%  Figures for Dan %%%%%%%%%%%%%%%%%%%%%%%%%%

Gribov proposed that the functional integral with respect to $A$ should extend only over what is now called the Gribov region \cite{Gribov:1977wm}.  It is the region where all eigenvalues of the Faddeev-Popov operator
\beq
M^{ac}(gA) = - D_i^{ac}(gA) \p_i - \p_i^2 \delta^{ac} - f^{abc} gA_i \p_i,
\eeq
are positive.  It was subsequently found that the eigenvalues are given by \cite{Zwanziger:1989mf}
\beq
\label{eigenvalue}
\lambda(\vec k; gA) = \vec k^2  \left( 1 - {H(gA) \over (N^2-1)dV } \right) + {\rm higher-order \ in} \ \vec k,
\eeq
where\footnote{For reviews see \cite{Sobreira:2004, Vandersickel:2012}}
\beq
\label{horizonfunction}
H(gA) = \int d^dx d^dy D_\mu^a(x) D_\mu^b(y) (M^{-1})^{ab}(x, y; A)
\eeq
is called the ``horizon function."  The eigenvalues $\lambda(\vec k; gA)$ are continuous functionals of $gA$.  At $gA = 0$ the eigenvalues are given by $\lambda(\vec k; 0) = \vec k^2$, where $\vec k_i = 2\pi n_i/L$, and $V = L^d$ is the quantization volume, and the $n_i$ are integers.  Here $\vec k$ serves as a label for the eigenvalue $\lambda(\vec k; gA)$ that is continuously connected to $\lambda(\vec k; 0) = \vec k^2$.  In the limit of large volume $L^d$, the spectrum approaches a continuum, and at small $\vec k$ the terms of higher order in $\vec k$ become negligible.  There is a common factor of $\vec k^2$ in~\eqref{eigenvalue}, so in the limit of large quantization volume an infinite number of eigenvalues pass through 0 together and become negative together, as illustrated in Fig.\ 1.  We call this phenomenon "all horizons are one horizon."  This is illustrated in Fig.\  2.  Gribov regions are sometimes represented as in the left side of Fig.\ 2, with distinct horizons for different eigenvalues.\footnote{Each zero of a every eigenvalue defines a Gribov horizon.}  This is correct at finite volume $L^d$.  At large volume, an infinite number of horizons coincide, as in the right side of Fig.\ 2.

%%%%%%%%%%%%%%%%%%%%%%%%%%%%%%%%%%%%%%%%%%%%%%%%%%%  Figures for Dan %%%%%%%%%%%%%%%%%%%%%%%%%%
\begin{figure}
        \begin{center}
                \includegraphics[width=7cm,clip]{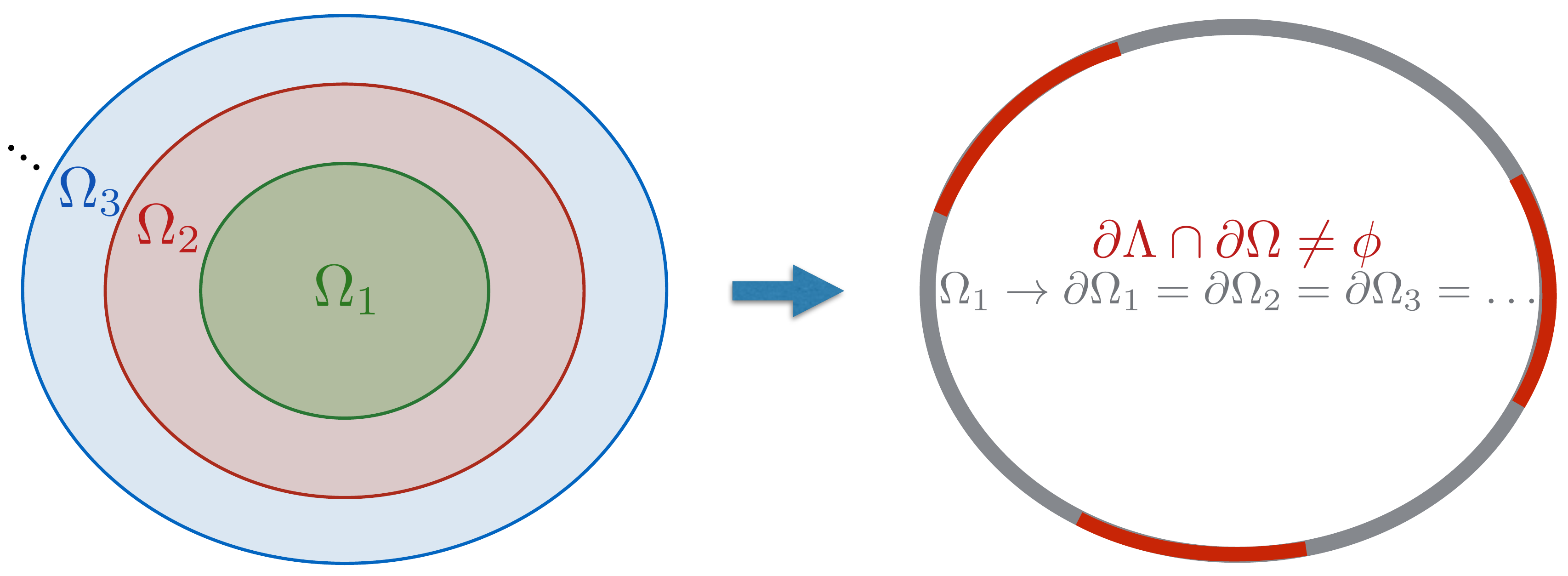}
                \caption{All horizons are one horizon}
                \label{one_horizon}
        \end{center}
\end{figure}
%%%%%%%%%%%%%%%%%%%%%%%%%%%%%%%%%%%%%%%%%%%%%%%%%%%  Figures for Dan %%%%%%%%%%%%%%%%%%%%%%%%%%

\section{Finite Temperature}

So far we have been discussing zero temperature.  At finite temperature $T$, a different picture emerges.  In the Euclidean formulation, finite temperature corresponds to finite period $1/T$ in the thermal direction, with discrete Matsubara frequencies $2\pi n T$.  In contrast, at large spatial volume the spectrum of the Faddeev-Popov operator for each $n$ approaches a continuous spectrum, as illustrated in Fig.\ 3.  As a result the horizon condition takes the form \cite{Cooper:1512.08}
\beq
\left\langle \int d^Dx d^Dy D_i^{ab} D_i^{ac}(y) (M^{-1})^{bc}(x, y) \right\rangle = V d(N^2-1),
\eeq
where $M = - D_\mu \p_\mu$,  $D = d + 1$, $\mu = 1,... D$, and $i = 1,... d$.

%%%%%%%%%%%%%%%%%%%%%%%%%%%%%%%%%%%%%%%%%%%%%%%%%%%  Figures for Dan %%%%%%%%%%%%%%%%%%%%%%%%%%
\begin{figure}
        \begin{center}
                \includegraphics[width=7cm,clip]{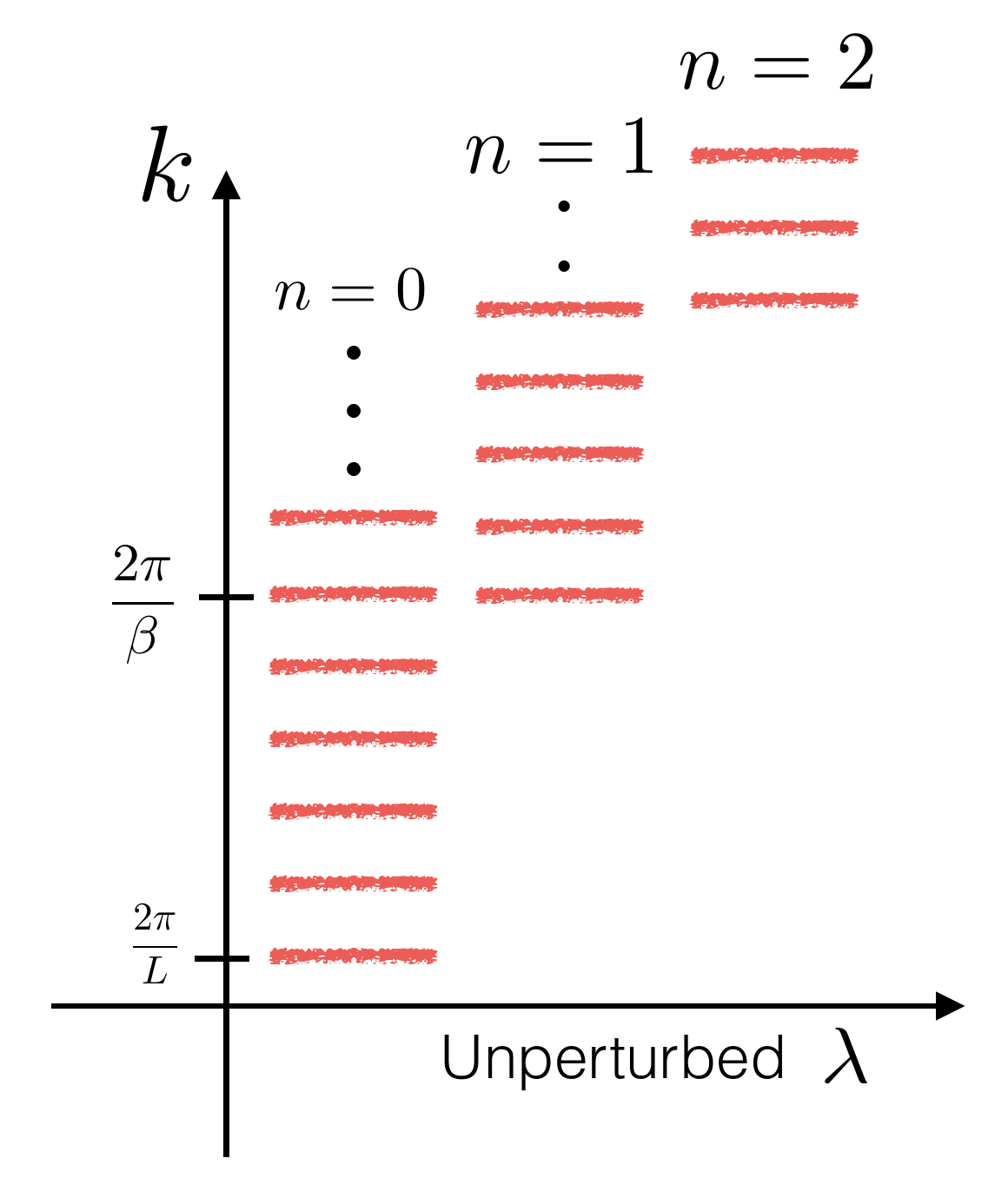}
                \caption{A sketch of the spectrum of the unperturbed
                        Faddeev-Popov operator in Landau gauge $M(0) = -
                        \p_\mu^2$ at finite temperature.  The index $n$ labels
                        the finite Matsubara frequencies.  When the
                        perturbation is turned on, $M(0) \to M(gA)$, the
                        eigenvalues corresponding to the zero-Matsubara
                frequency cross the Gribov horizon first.}
                \label{levels}
        \end{center}
\end{figure}
%%%%%%%%%%%%%%%%%%%%%%%%%%%%%%%%%%%%%%%%%%%%%%%%%%%  Figures for Dan %%%%%%%%%%%%%%%%%%%%%%%%%%

\section{Horizon Function and Non-local Action}

We return to zero temperature.  The Euclidean action is given by,
\beq
S = S_{\rm FP} + \gamma H - \gamma \int d^Dx \ (N^2 - 1) D,
\eeq
where $S_{\rm FP}$ is the Faddeev-Popov action, and the horizon function $H$ is given in~\eqref{horizonfunction}.

The horizon function cuts off the functional integral at the Gribov horizon, as one sees from the eigenfunction expansion
\beq
(M^{-1})^{ab} = \sum_n { \psi_n^a(x) \psi_n^b(y) \over \lambda_n(gA) }.
\eeq

\section{Horizon Condition and Kugo-Ojima Confinement Condition}

The Gribov parameter is fixed by the horizon condition,
\beq
\langle H \rangle = (N^2-1)d \int d^dx.
\eeq
It is a remarkable fact that the horizon condition and the famous Kugo-Ojima confinement condition are the same statement \cite{Kugo-Ojima, Kugo:1995km}
\beq
-i \int d^dx \ \left\langle (D_\mu c)^a(x) (D_\mu \bar c)^a(0) \right\rangle = (N^2-1)d.
\eeq
This may indicate that color confinement is assured in this theory, although the precise hypotheses of the Kugo-Ojima theorem are not satisfied in this approach.

\section{Horizon Condition and Dual Meissner effect}

It is a remarkable fact that the horizon condition is equivalent to the statement that the QCD vacuum is a perfect color-electric superconductor, which is the dual Meissner effect \cite{Reinhardt:2008},
\beq
G(\vec k) = {d(\vec k) \over \vec k^2 } = {1 \over \epsilon(\vec k) \vec k^2 }
\eeq
\beq
d^{-1}(\vec k = 0) = 0 \Longleftrightarrow \epsilon(\vec k = 0),
\eeq
where $G(\vec k)$ is the ghost propagator and $\epsilon(\vec k)$ is the dielectric constant.

\section{Auxiliary ghosts}

Just as the Faddeev-Popov determinant is localized by introducing ghosts,
\beq
\det M = \int dc d \bar c \exp\left( - \int d^dx \ \bar c M c \right),
\eeq
likewise, the horizon function in the action may be localized by introducing ``auxiliary" ghosts \cite{Zwanziger:1989mf},
\beq
\exp( - \gamma H) = \int d\varphi d \bar\varphi d \omega d \bar\omega \exp\left(- \int d^dx \  \left[ \bar\varphi M \varphi - \bar\omega M \omega + \gamma^{1/2} D \cdot (\varphi - \bar\varphi ) \right] \right).
\eeq

\section{Local Action and Physical Degrees of Freedom in Coulomb Gauge}

In terms of the additional fields, the local action is given by \cite{Cooper:1512.05},
\beqa
S & = & \int d^{d+1}x \ \{ i \tau_i D_0 A_i + (1/2)[\tau_i^2 + (\p_i \lambda)^2 + (1/4)F_{ij}^2 
+ i \p_i \lambda D_i A_0 - \p_i \bar c D_i c  +   \p_i \bar\varphi_j \cdot D_i \varphi_j 
\nonumber \\
& & \ \ \ \ \ \ \
 - \  \p_i \bar\omega_j \cdot ( D_i \omega_j + D_i c \times \varphi_j) 
 + \gamma^{1/2} {\rm Tr} [ D_j(\varphi - \bar\varphi)_j - D_j c \times \bar\omega_j ] - \gamma (N^2 -1) d \},
\eeqa
Here we have introduced the canonically conjugate color-electric field $\pi$, and decomposed it into its transverse and longitudinal parts
\beq
\pi_i = \tau_i + \p_i \lambda,
\eeq 
where $\tau_i$ is the momentum conjugate to $A_i$, and both are transverse,
\beq
\p_i \tau_i = \p_i A_i  = 0.
\eeq
Only the first term,
\beq
i\tau_i D_0 A_i = i \tau_i \p_0 A_i + i \tau_i A_0 \times A_i,
\eeq
contains a time derivative.  It acts on the two transverse degree of of freedom of the gluon which are the would-be physical degrees of freedom.  All the remaining terms impose constraints.

\section{Maggiore-Schaden Shift}

It is another remarkable fact that the same action may be derived by a completely different method developed by Maggiore and Schaden, without reference to the Gribov horizon \cite{Schaden:1994, Schaden:1996}.

Start with the usual Faddeev-Popov fields and action, on which the BRST operator $s$ acts according to
\beqa 
sA_\mu = D_\mu c \hspace{3cm} sc  =   - (g/2)(c \times c) 
\nonumber \\
s \hat{\bar c} = i \hat b \ \ \    \hspace{3cm} s \hat b = 0 \ \ \ \  \ \ \ \  \  \ \ \ \ \ \ \ 
\nonumber   \\
s \pi_i = g\pi_i \times c, \hspace{5.6cm} ,
\eeqa
Introduce auxiliary quartets of bose and fermi ghosts whose determinants cancel when they are integrated out, on which $s$ acts according to
\beqa
s \phi_B = \omega_B  \hspace{3cm} s \omega_B = 0 \
\nonumber  \\
\ \  s \bar\omega_B = \bar\phi_B  \hspace{3cm} s \bar\phi_B = 0.
\eeqa
The resulting action is BRST-invariant by construction,
\beq
\mathcal L = \mathcal L^{\rm YM} + \mathcal L^{\rm gf} = \mathcal L^{\rm YM} + s \Psi  
\hspace{3cm} s \mathcal L = 0,
\eeq
where $\mathcal L^{\rm FP} = (1/4) F_{\mu \nu}^2$ and
\beq
\Psi \equiv \p_i \hat{\bar c} \cdot A_i  + \p_i \bar\omega_j \cdot D_i \phi_j
\eeq
\beq
s \Psi = i \p_i \hat b \cdot A_i  - \p_i \hat{\bar c}_j \cdot D_i c + \p_i \bar\phi_j \cdot D_i \phi_j - \p_i \bar\omega_j \cdot ( D_i \omega_j D_i c \times \phi_j).
\eeq

Now make the change of variable
\beqa
\phi_{jb}^a(x) & = & \varphi_{jb}^a(x) - \gamma^{1/2} x_j \delta_b^a
\nonumber \\
\bar\phi_{jb}^a(x) & = & \bar\varphi_{jb}^a(x) + \gamma^{1/2} x_j \delta_b^a
\nonumber \\
\hat{\bar c}^a(x) & = & \bar c^a(x) - \gamma^{1/2} x_j f^{abc} \bar\omega_{jc}^b(x)
\nonumber \\
\hat b^a(x) & = & b^a(x) + i\gamma^{1/2} x_j f^{abc} \bar\varphi_{jc}^b(x).
\eeqa
This procedure, by a completely different derivation, yields precisely the same $x$-independent Lagrangian density that was derived from a cut-off at the Gribov horizon.  The resulting lagrangian density remains BRST-invariant, $s \mathcal L = 0$, but the {\em form} of the BRST operator $s$ acting on the new fields is changed.  (For an alternative approach and further references, see \cite{Capri:2016}.)

\section{Spontaneous breaking of BRST}

The BRST operator acts on all the shifted fields exactly as it does on the corresponding unshifted fields, except for
\beq
s \bar\omega_{jb}^a = \bar\varphi_{jb}^a + \gamma^{1/2} x_j \delta_b^a.
\eeq 
We have regained an $s$-invariant, local, $x$-independent lagrangian density, but the BRST symmetry is spontaneously broken
\beq
\langle \{ Q_B, \bar\omega_{ib}^a \} \rangle = \langle s \bar\omega_{ib}^a \rangle = \gamma^{1/2} x_i \delta_b^a.
\eeq

When BRST is unbroken, physical states are characterized as being invariant under the action of the BRST operator, $Q_B \Phi_{\rm phys} = 0$ (and we mod out the states of norm zero).  The last equation implies that the vacuum state is not BRST-invariant,\footnote{This implies that physical states cannot be defined as the cohomology of $s$.} 
\beq
\langle \{ Q_B, \bar\omega_{ib}^a \} \rangle = (Q_B \Phi_0,  \bar\omega_{ib}^a \Phi_0) + (\Phi_0,  \bar\omega_{ib}^a Q_b \Phi_0) = \gamma^{1/2} x_i \delta_b^a \neq 0,
\eeq
so we require another characterization of physical states.

\section{Physical states in Faddeev-Popov Theory}

For purposes of orientation, let us review quantization of Faddeev-Popov theory.  The Faddeev-Popov determinant is non-local.  It is localized at the cost of introducing the Faddeev-Popov ghost which creates unphysical states. There is also a benefit however, because the local action $\mathcal L = \mathcal L^{\rm YM} + s \Psi$, has a new symmetry, the BRST-symmetry, with $s \mathcal L = 0$, that is not present in the non-local theory.  This symmetry allows us to characterize physical observables: they are the operators $G$ that are invariant under the BRST symmetry,
\beq
\mathcal W_{\rm phys} \equiv \{ G: sG = 0 \}.
\eeq
Physical states are obtained by applying physical operators to the vacuum state.

\section{Physical states in the GZ theory}

Localization of the cut-off at the Gribov horizon yields
\beq
\mathcal L = \mathcal L^{\rm YM} + s \chi = \mathcal L^{\rm YM} + \mathcal L_{\rm aux.gh}.
\eeq
where $\mathcal L_{\rm aux.gh}$ now depends not only on  the Faddeev-Popov ghosts, but also on the auxiliary ghosts.  This action possesses many new symmetries, with generators $Q_Y$ that leave ordinary physical observables invariant,
\beq
[ Q_Y, \mathcal L] = [ Q_Y, F^2 ] = [Q_Y, \bar \psi \psi ] = 0.
\eeq
A classification of the these symmetries may be found in \cite{Schaden:1412, Schaden:1501}.  We call them ``phantom symmetries" because they act on these physical operators like the zero-operator, but they act non-trivially on the auxiliary ghost fields.  It is natural to define physical observables as the class of operators that are invariant under all the new phantom symmetries \cite{Schaden:1412},
\beq
\mathcal W_{\rm phys} \equiv \{ G: sG = 0; [Q_Y, G] = 0 \} {\rm \ for \ all} \ Q_Y.
\eeq

We offer as a {\em conjecture} that BRST symmetry is not broken by $s$-exact operators in $\mathcal W_{\rm phys}$,
\beq
\langle sY \rangle = 0 \ {\rm for \ all} \ sY \in \mathcal W_{\rm phys}.
\eeq
In other words, BRST symmetry is preserved {\em where it is needed}, namely in the physical subspace.

Explicit calculation of several examples reveals that BRST symmetry-breaking apparently afflicts the unphysical sector, but may be unbroken where it is needed, namely in cases of physical interest.  For example, the BRST-exact part of the conserved energy-momentum tensor has vanishing expectation-value,
\beq
T_{\mu \nu} = T_{\mu \nu}^{\rm YM} + s \Xi_{\mu \nu}
\eeq
\beq
\langle s \Xi_{\mu \nu} \rangle = 0.
\eeq
As before physical states are obtained by applying physical operators to the vacuum state.

\section{Conclusion}
We have a local, renormalizable quantum field theory with the following interesting properties:\\
$\bullet$  It provides a cut-off at the Gribov horizon.\\
$\bullet$ The Kugo-Ojima color confinement condition is satisfied.\\
$\bullet$ The vacuum is a perfect dielectric.\\
$\bullet$ There is an alternate derivation by the Maggiore-Schaden shift that provides a BRST-invariant Lagangian.\\
$\bullet$ BRST-symmetry is spontaneously broken, but perhaps only in the unphysical sector.\\
$\bullet$ In Coulomb gauge, the color-Coulomb potential is linear or super-linear when the Wilson potential is linearly rising.

\end{document}